# Impact of pandemic fatigue on the spread of COVID-19: a mathematical modelling study

Disheng Tang[1,2], Wei Cao[3*], Jiang Bian[3], Tie-Yan Liu[3], Zhifeng Gao[3], Shun Zheng[3], Jue Liu[4*]

[1]Interdisciplinary Graduate Program in Quantitative Biosciences, Georgia Institute of Technology, Atlanta, GA, USA. [2]School of Earth and Atmospheric Sciences, Georgia Institute of Technology, Atlanta, GA, USA. [3]Microsoft Research Asia, Beijing, China. [4]Department of Epidemiology and Biostatistics, School of Public Health, Peking University, Beijing, China. [*]wei.cao@microsoft.com; jueliu@bjmu.edu.cn.

**Abstract**
In late-2020, many countries around the world faced another surge in number of confirmed cases of COVID-19, including United Kingdom, Canada, Brazil, United States, etc., which resulted in a large nationwide and even worldwide wave. While there have been indications that precaution fatigue could be a key factor, no scientific evidence has been provided so far. We used a stochastic metapopulation model with a hierarchical structure and fitted the model to the positive cases in the US from the start of outbreak to the end of 2020. We incorporated non-pharmaceutical interventions (NPIs) into this model by assuming that the precaution strength grows with positive cases and studied two types of pandemic fatigue. We found that people in most states and in the whole US respond to the outbreak in a sublinear manner (with exponent k=0.5), while only three states (Massachusetts, New York and New Jersey) have linear reaction (k=1). Case fatigue (decline in people's vigilance to positive cases) is responsible for 58% of cases, while precaution fatigue (decay of maximal fraction of vigilant group) accounts for 26% cases. If there were no pandemic fatigue (no case fatigue and no precaution fatigue), total positive cases would have reduced by 68% on average. Our study shows that pandemic fatigue is the major cause of the worsening situation of COVID-19 in United States. Reduced vigilance is responsible for most positive cases, and higher mortality rate tends to push local people to react to the outbreak faster and maintain vigilant for longer time.

As warned by researchers (1), the resurgence of COVID-19 happened around the world in late 2020 (2,3,4). Some countries like Italy and Iran had a second wave earlier than most (5,6), while others such as United States, United Kingdom and Canada experienced record-breaking cases (7). Take United States for example, as of Dec 22, cases of the past two months already account for more than half (53%) of total cases in the whole 2020. Even though there has been a significant decrease in COVID cases since 2021, it is important to understand the mechanisms of the multiple waves and prevent potential shortages of equipment, staff and beds in hospitals (8).

However, the principles behind the resurgence of COVID-19 remain a challenging problem. While many studies on second wave focus on the optimal strategy for school opening or control measures (9,10), compartment models with homogeneous mixing assumption, and other methods such as network and spatial models usually explore the additional epidemic states (11,12) or the interaction structure (13) and mostly overlook the resurgence. Although many public health experts have expressed their concerns that pandemic fatigue could threaten global health and might be the reason of the surge (14,15,16), few studies focused on quantifying its impact on the spread of coronavirus (17).

Here we adopted the hierarchical metapopulation model that has shown capable of characterizing recurrent outbreaks and simulating real epidemics such as Severe Acute Respiratory Syndrome (SARS) (18). We incorporated into it a precaution strength that varies based on recent positive cases and decays with time due to pandemic fatigue and used the cases in United States as an example to conduct our modeling study. We examined the mechanisms of the COVID-19 resurgence, and the explanations for its increasing epidemic size. We explored the situations without pandemic fatigue, hoping to provide insights for precaution strategy.

We fitted our model to the whole curve of daily positive cases of 22 states and the whole US cases between March 4th and December 22nd, 2020, 7-day moving average is used to filter out the noise. Here we showed the results of US and its five states and put the rest in the Supplementary Appendix. We ran 5000 trials using hyperparameter optimization algorithm to find the optimal parameters. We found that the optimal memory length and speed of response $k$ found by optimization algorithm are correlated with death rate in each state (Figure S1 in Supplementary Appendix), indicating that people in places with higher mortality rate tend to react faster, and remember older outbreak(s) thus take NPIs for a longer time. As a result, there are fewer waves in New York, New Jersey and Massachusetts where mortality rate is much higher than the rest. We also found that there's a strong negative correlation between the time-varying reproduction number and the fraction of vigilant group. This relation is true for all data sets (the rest in Supplementary Appendix), and they even have close identical slopes with slightly different intercepts. This indicates that the fraction of individuals taking NPIs controls the reproduction number and determines specific features of the outbreak, including the size and number of waves.

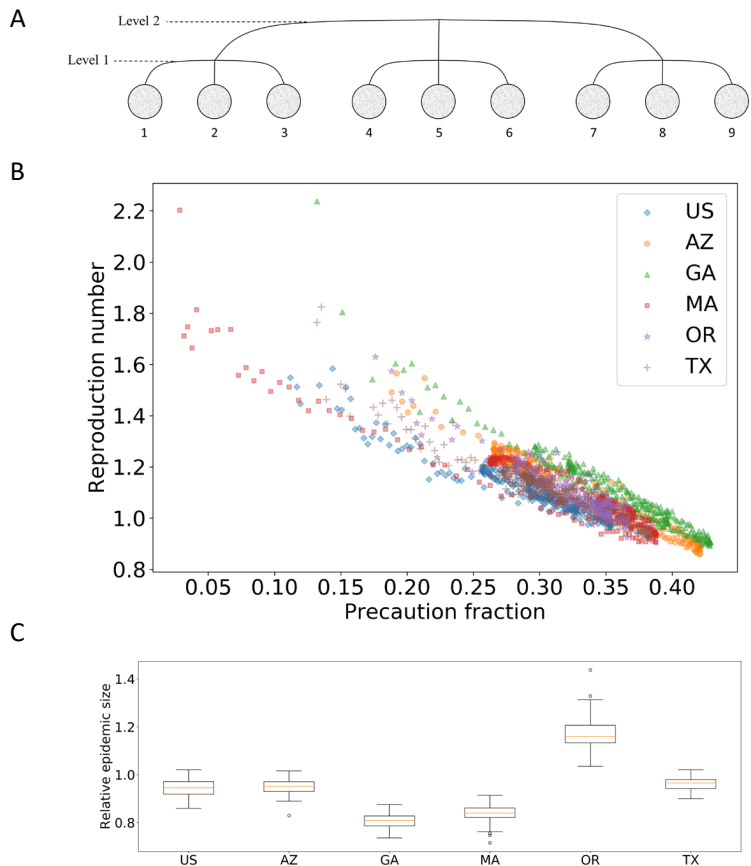

*Figure 1: Schematic of the hierarchy structure and features of model fitting*
(A) An example of hierarchical structure with branching ratio $b = 3$ and total depth $l = 2$. There are $b^l = 9$ contexts (circles) in this example, and within each context the mean field assumption holds true. An individual travels from context $i$ to $j$ with probability $q_{ij} \propto e^{-\frac{d_{ij}}{\xi}}$, where $d_{ij}$ is the ultra-metric distance between contexts $i$ and $j$ and $\xi$ is a tunable parameter. $d_{ij}$ is defined as the lowest level the subpopulations $i$ and $j$ share. For example, here $d_{12} = 1, d_{24} = d_{68} = 2$. (B) Dynamic reproduction number against fraction of vigilant group for six example data sets: the whole United States, Arizona, Georgia, Massachusetts, Oregon and Texas states. (C) Relative epidemic size of 100 model simulations over real data.

As Figure 1 shows, the epidemic size generated from our model is close to the real case, since most boxes are centered around a relative ratio of 1. Even the single simulation of Oregon State with the largest ratio of 1.44 only has an epidemic size of 3.5%, leaving 96.5% of uninfected population. The ratio distributions for the remaining are all within the range of [0.4, 1.5] (Figure S3 in Supplementary Appendix). Because of the relatively close size, our model can simulate a complete estimated dynamical process and predict a long-term run of coronavirus outbreak with higher reliability, in contrary to traditional epidemic models in which the simulated size is usually much larger than real scenario and the pandemic dies down fast mostly due to the depletion of susceptible individuals.

Due to the stochasticity of this model, we ran 100 simulations for each optimal parameter setting. We found that $k = 1$ is true only for Massachusetts, New York and New Jersey States, and $k = 0.5$ for the remaining. This means that only these three states react to COVID-19 in a linear manner, while the response of others is much slower. Figure 2 shows the small variations in different runs, while the average is quite close to the real case. The largest relative fitting error (New York in appendix) is 40%, while the smallest (the whole US) is 11%. It is obvious that the states with longer memory have smoother variations of vigilant group, such as Massachusetts State with a 170-day memory. The long-haul fight against the coronavirus directly leads to the absence of a second wave around August in these states, reducing potentially a considerable amount of COVID cases. The growth in vigilant group is always several weeks later than positive cases because of the averaging and sublinear response in many states. It is worth noticing that the vigilance fraction $v(t)$ already increased to nearly its maximal value $f(t)$ in response to the first and smallest outbreak, however due to case fatigue, a 10 times larger follow-up wave cannot even alarm an equal number of people! This indicates that case fatigue probably takes most responsibility for the record-breaking coronavirus wave in the US in late 2020.

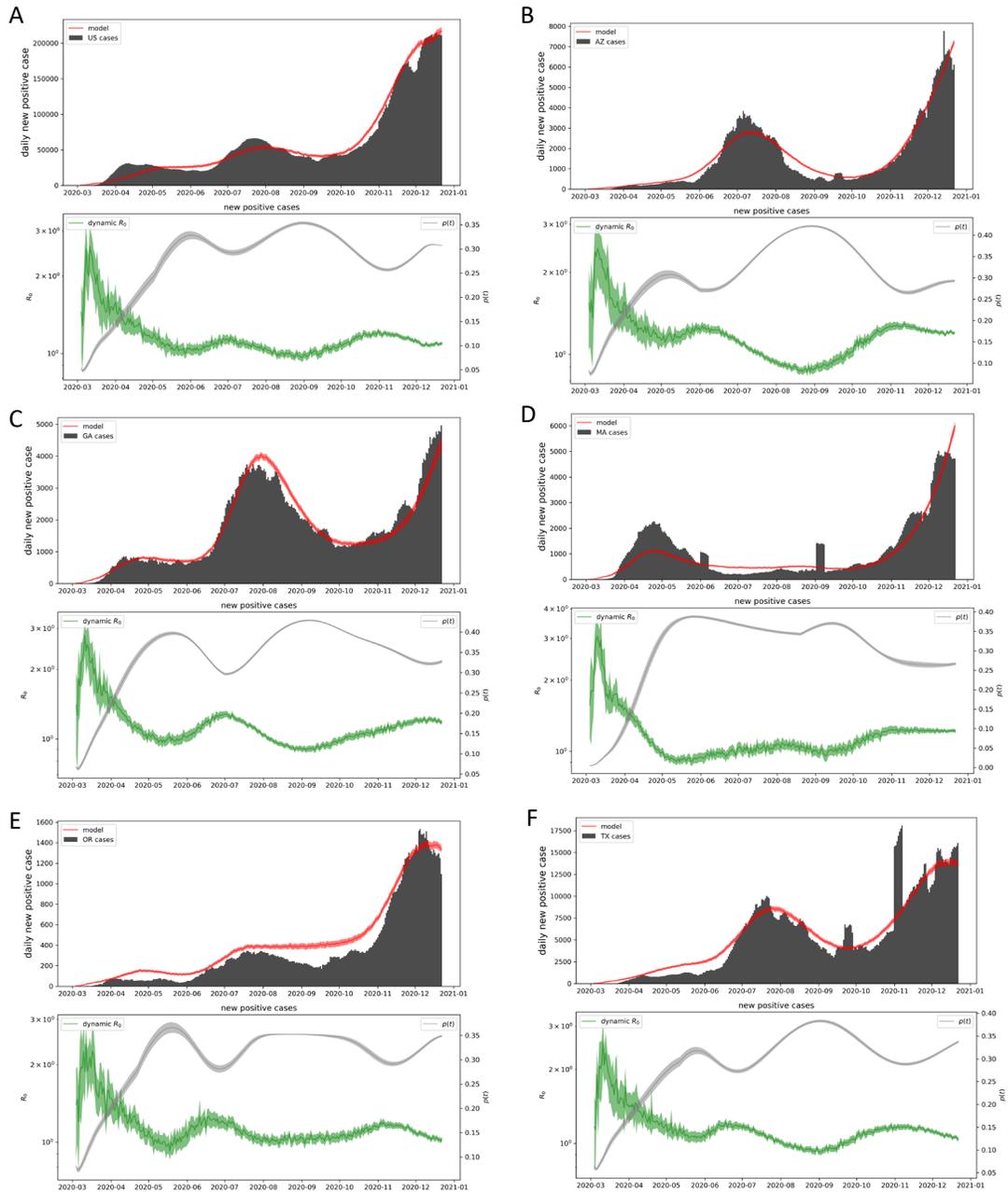

*Figure 2: Model fit of six data sets with a hierarchically structured population with branching ratio $b = 6$, depth $l = 5$, group size $n = 100$, transmission rate $\beta = 0.65$ and recovery rate $\gamma = 0.2$.*
(A-F) Top panels are the fitting results, where bars correspond to the real data of the whole United States, Arizona, Georgia, Massachusetts, Oregon and Texas respectively, red line is the average of 100 simulations with its 99% confidence interval (CI). The bottom panels are the time-varying reproduction number and the vigilance fraction $p(t)$ with their 99% CIs.

In Figures S7 and S9, we showed that our model has some good predictive power: when predicting the cases of following 15 days, its smallest error (the whole US) is merely 3.8%, while the largest (Wyoming) is 56% and the average error is 23%. For predicting a longer future like the next 30 days, our model has a larger average error of 75%. However, our prediction of the time of future outbreaks is still satisfactory, which could provide valuable warning for potential waves.

To study the impact of pandemic fatigue on the spread of coronavirus, we explored the situations where memory length is longer and precaution fatigue (decay in maximal vigilance fraction $f(t)$) or case fatigue (decline in case-sensitivity $c(t)$) is smaller and even absent. Surprisingly, we found that the extension of memory would not lead to a decline in epidemic size; on the contrary, it generated an increase in affected number of individuals when memory length was larger enough to be considered as unlimited (see Figure 3 (G)). This indicated that longer memory is not always good-- when people remember waves long time ago, averaging effect tend to make them think the upcoming wave is nothing terrible.

Nonetheless, we found both precaution and case fatigue are major reasons of the exceptionally large size of COVID-19. With the increase of both lower bounds $l_f$ and $l_c$, final epidemic size declined quite fast, especially those data sets with extremely small original values, e.g., Arizona State in Figure 3 (I). Figure 3 (J) also showed that when there was no pandemic fatigue ($l_f = l_c = 1$), five out of six data sets would have a more than 70% decrease in epidemic size. In Figure 3 (A-F) we can see that all data sets would have had a much smaller outbreak if there were no decay in precaution strength, and the occurrence of COVID fatigue is distinct in each scenario. In fact, of all 23 data sets used in this study, pandemic fatigue accounts for 68% of total coronavirus cases on average. This indicates that COVID fatigue is the largest cause of the worsening coronavirus situation in late 2020.

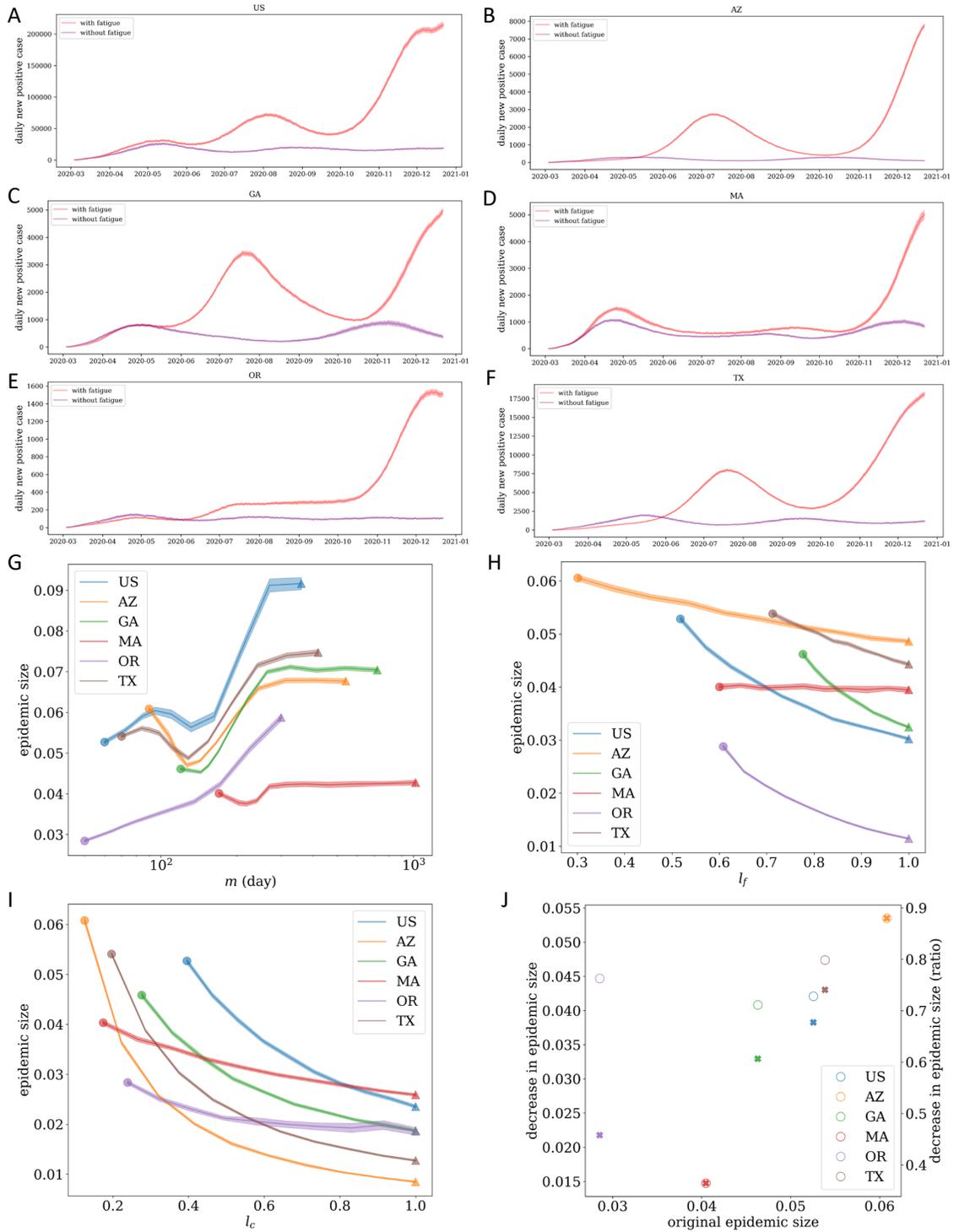

**Figure 3: Impact of pandemic fatigue on the outbreak**
(A-F) Original cases with pandemic fatigue and cases without pandemic fatigue for six data sets. Both curves are plotted with 99% confidence interval. (G) Epidemic size against memory length $m$ with 99% confidence interval. Circles represent the original result, and triangles denote the situation where the memory is unlimited. The value of $m$ at the triangle is larger than the observation length for all data, thus can be regarded as unlimited memory. (H, I) Epidemic size against the lower bounds $l_f$ and $l_c$ with 99% confidence interval. Circles are the original results, and triangles show the situations where there were no precaution and case fatigue, respectively. (J) The decrease and decrease ratio in epidemic size when there is no pandemic fatigue, i.e., $l_f = l_c = 1$. "X" signs correspond to the absolute value of decrease while empty circles represent the decrease in ratio.

We incorporated NPIs into a hierarchical metapopulation model and considered two types of pandemic fatigue: precaution and case fatigue. Using United States for example, we obtained fairly good fits of the whole curve of positive cases between March and December 2020. We also found that the model could predict the cases of several states in the next 15 days with good accuracy, and even though it overestimated the cases in the next 30 days for some states, it was still able to predict the right timing of the upcoming wave, providing valuable warnings for a long time ahead.

We found that the estimated reproduction number had a perfect negative correlation with the fraction of individuals seeking to take NPIs. Although some states had different intercepts, they all had close slopes in this correlation, meaning the variation in vigilant group had similar impact in each case. This finding suggested that one effective control measure is to encourage people to take precaution measures. The decline of early waves was caused by stronger precaution measures, and we found that the emergence of larger follow-up waves originated from the decrease in vigilant group and decay in vigilance level, which we interpreted as two signs of pandemic fatigue.

While other research focused on the fatigue of policy makers and health workers (19,20), our study explored civilian-level COVID fatigue and suggested it is one of the major reasons behind the resurgence of coronavirus. Although there has been significant decrease in COVID-19 cases since 2021 due to effective preventive interventions such as vaccines, pandemic fatigue will only intensify in the near future hence needs our attention to prevent potential resurgence.

However, there are some limitations to our model analysis. We assumed that the precaution strength only decreases with time in the long-term, while there could be some temporary increase in the interim due to stronger policies or higher productivity. For example, in many places personal protective equipment (PPE) such as facemask, respirator and hand sanitizer were in extreme shortage during the early outbreak; but as manufacturers continued to produce the equipment, more and more people gained access to PPE which might result in a quick decrease in cases and a long period with the disease under control. This could possibly be one major reason that our model fitted the cases in New York and New Jersey states with much higher error than others.

Our model focused on the effect of NPIs in analysis of COVID-19, but didn't explore the impact of travel restrictions. One of the major advantages of hierarchical metapopulation model is its contexts located in a hierarchical structure with travelers to spread the disease from one to another, while mean-field assumption is true within each subpopulation. This indicates that one possible strategy to control the spread of pandemic influenza is to impose travel ban and keep the epidemic local. When most infected or susceptible individuals got into the recovered state, the disease simply dies down. We merely used constant values for both the number of expected infected travelers $P_0$ and travel distance tendency $\xi$ due to high complexity, however, there are likely to be better implementations of the model with higher efficiency or just more computing power that might make it possible to study time-varying travel parameters.

Even though our model could fit the real cases well, we used only three states (susceptible, infected, recovered) for the transmission model. There have been many studies using SEIR model or even more compartments (21,22,11) and showed that their epidemic models can also have good fits. This suggests that more compartments are worth trying if our model failed to fit more complicated curves of future cases.

In conclusion, our results show that COVID fatigue is a major reason behind the resurgence in coronavirus cases. In order to prevent another outbreak of COVID-19 cases, we need to maintain vigilant and overcome pandemic fatigue. This long-term war against SARS-CoV-2 is not over yet, we can never let our guard down.

## Methods

Transmission model and precaution with fatigue are the two main components of our model. We used SIR model with a hierarchical structure as the transmission model, assuming that homogeneous mixing is true withing each subpopulation, and individuals travel between contexts with a probability correlated with their distance. We introduced a precaution strength that is only related to recent positive cases due to underestimation of the disease, and gradually decays with time because of COVID fatigue. We studied pandemic fatigue through caution and case fatigue, representing the decline in number of vigilant people and individuals becoming numb to new confirmed cases. In order to test the predictive power of our model, we partitioned the real cases into training set and test set, and ran 5000 trials to find the best fit on the training data.

**Transmission model**

To consider the spatial spread of COVID-19, we used a stochastic metapopulation model with subpopulations distributed in a hierarchical structure (18). Each subpopulation equally consists of $n$ individuals, while the branching ratio $b$ and total depth $l$ together decide the number of groups ($b^l$). Therefore, the total population of this model is $N = b^l \times n$. Figure 1 (A) shows an example of the hierarchical structure.

All individuals except for one start as susceptible to SARS-CoV-2 infection, and the one initial infected case is assigned to a random subpopulation. Within each local context, compartment model (SIR) with transmission rate $\beta$ and recovery rate $\gamma$ is adopted. At each timestep, each individual travels from context $i$ with probability $p$ and enters a new context $j$ with probability $q_{ij} \propto e^{-\frac{d_{ij}}{\xi}}$, where $d_{ij}$ is the ultra-metric distance between contexts $i$ and $j$, while $\xi$ is a tunable parameter that controls the tendency of long-distance traveling. Defined as the expected number of infected individuals leaving a single context over the mean infectious period, $P_0$ basically controls how many potential spreaders there are and it has been shown by Watts et al. (18) that $P_0 \approx p\psi'n/\gamma$, where $\psi'$ is the expected normalized epidemic size for a compartment model with reproduction number $R_0$. In our model we used $\beta = 0.65, \gamma = 0.2, b = 6, l = 5, n = 100$ which constitute a total of 777600 individuals for all data set. Through our experiments we found that $b$ and $l$ need to be large enough when fitting complicated spreading process of infectious disease and showed in the results that our parameter settings enabled our model to simulate any scenario we encountered.

**Precaution with fatigue**

In order to incorporate the non-pharmaceutical interventions (NPIs) and COVID fatigue into this model, we considered a group of vigilant individuals that seek to take precaution measures based on recent confirmed COVID-19 cases, and assumed there are two types of fatigue, of which the first one is precaution fatigue denoting less vigilant people and the second is case fatigue meaning lower vigilance level to new cases. We defined the vigilant group below

$$v(t) = f(t) \min\left(\left[c(t)\frac{\sum_{\tau=t-(m-1)}^{t} \dot{I}(\tau)}{mN}\right]^k, 1\right)$$

where $f(t)$ represents the maximal fraction of vigilant group at time $t$, $c(t)$ is the case-sensitive factor at time $t$, $\dot{I}(t)$ is the new positive cases at time $t$ and $k$ controls how fast people react to new cases (sublinear, linear or superlinear). Since interventions are usually adjusted regarding the epidemic situations over a period, we used the average of new cases for $m$ time steps to manage the intervention strength, where $m$ represents the memory length. When there's an outbreak, the surge in $\dot{I}(t)$ will lead to an increase in $v(t)$, i.e., more people taking NPIs which makes the disease harder to spread. As the epidemic eases people start to let their guards down and this is usually when a second wave attacks. It's obvious that if $f(t)$ and $c(t)$ decline, there will be less vigilant people and they will become less sensitive to positive cases, leading to fierce resurgence of cases. Therefore, $f(t)$ corresponds to precaution fatigue while $c(t)$ shows case fatigue.

We used reverse sigmoid functions both for $f(t)$ and $c(t)$ to emulate the phenomenon where fatigue permeates slowly

$$f(t) = f(0) \frac{l_f + e^{u_f - \frac{t}{s_f}}}{1 + e^{u_f - \frac{t}{s_f}}}$$

$$c(t) = c(0) \frac{l_c + e^{u_c - \frac{t}{s_c}}}{1 + e^{u_c - \frac{t}{s_c}}}$$

where $l_f$ and $l_c$ are the relative lower bounds of $f(t)$ and $c(t)$, $u_f, u_c$ and $s_f, s_c$ controls when and how fast these functions decay. In our model, pandemic fatigue (the decline in $f(t)$ and $c(t)$) is a major reason of the recent surge in cases. We showed the curves of $f(t)$ and $c(t)$ for each data set in Figure S4 in Supplementary Appendix.

Based on the paper by Eksin et al. (23), we assumed that the same fraction of both susceptible and infected individuals belongs to the vigilant group and take NPIs, and all of them cannot infect or get infected from others. Therefore, the dynamics become

$$\frac{dS}{dt} = -\frac{\beta(1-v(t))S(1-v(t))I}{N} = -\frac{\beta(1-v(t))^2 SI}{N}$$

where $v(t)$ is the fraction of vigilant group in the whole population.

**Data sources**

Since we used United States for an example in our modeling study, we fitted a stochastic transmission dynamic model to multiple publicly available data sets of positive cases in 22 states and the whole US. The data we used is from "The COVID Tracking Project" https://covidtracking.com/ and is open to public. The reason that we only used data of 22 states instead of all 50 in the United States is due to the incompleteness of the remaining data, considering we focused on the time period between March 4[th], 2020 and December 22[nd], 2020.

**Contributors**

Tang DS collect the data, did the analyses, produced the tables, and wrote the first draft of the manuscript. Cao W and Liu J conceived and designed the study. Bian J, Liu YT, Gao ZF, Zheng S, and Liu J revised the manuscript. All authors have read, contributed to, and approved the final version of the manuscript.

**Declaration of interests**

We declare no competing interests.

**Acknowledgments**

This work was supported by the National Key Research and Development Project (2020YFC0846300).

# Supplementary appendix for

# Impact of pandemic fatigue on the spread of COVID-19: a mathematical modelling study

Disheng Tang[1,2], Wei Cao[3]*, Jiang Bian[3], Tie-Yan Liu[3], Zhifeng Gao[3], Shun Zheng[3], Jue Liu[4]*

[1]Interdisciplinary Graduate Program in Quantitative Biosciences, Georgia Institute of Technology, Atlanta, GA, USA
[2]School of Earth and Atmospheric Sciences, Georgia Institute of Technology, Atlanta, GA, USA
[3]Microsoft Research Asia, Beijing, China
[4]Department of Epidemiology and Biostatistics, School of Public Health, Peking University, Beijing, China

*Correspondence to: wei.cao@microsoft.com, jueliu@bjmu.edu.cn

**Appendix Contents**

**Supplementary Text**

Data
The COVID-19 data sets we used were from "The COVID Tracking Project" (1), the population data we used was from World Population Review (2), we used positive cases and death cases from this data set (3)

Loss function
For model fitting and prediction, we used a commonly used hyperparameter optimization algorithm named Hyperopt (4) to find the optimal parameters. We defined a simple loss function as the relative absolute error

$$loss = \frac{\sum_{t=1}^{T}|Y_t - \hat{Y}_t|}{\sum_{t=1}^{T} Y_t}$$

$$\hat{Y}_t = \ddot{Y}_t \frac{\wp}{N}$$

where $T$ is the total length of data, $Y_t$ is the actual number of cases at time $t$, $\hat{Y}_t$ is the predicted number of cases at time $t$, $\ddot{Y}_t$ represents the number of model cases at time $t$, $\wp$ denotes the population of the data set and $N$ is the total number of individuals in the model. Since we fixed the hierarchy structure, during optimization we fitted the normalized data (cases/population), after we found the optimal parameters we adjusted the predicted cases by multiplying a constant $\frac{\wp}{N}$.

Hyperparameter space
For the expected number of infected travelers $P_0$ and long-distance travel tendency $\xi$, we used a suitable range to make sure the spreading speed and final epidemic size are both in a proper scope. It's worth noting that we set the range of $\xi$ as [0·4, 0·5], with the upper bound a little lower than the theoretical value $\frac{1}{\ln b}$ =0·558 that assumes people travel to all distances with equal likelihood (4), which is nearly impossible in real life. The lower bound of $\xi$ is to make sure there's an outbreak.

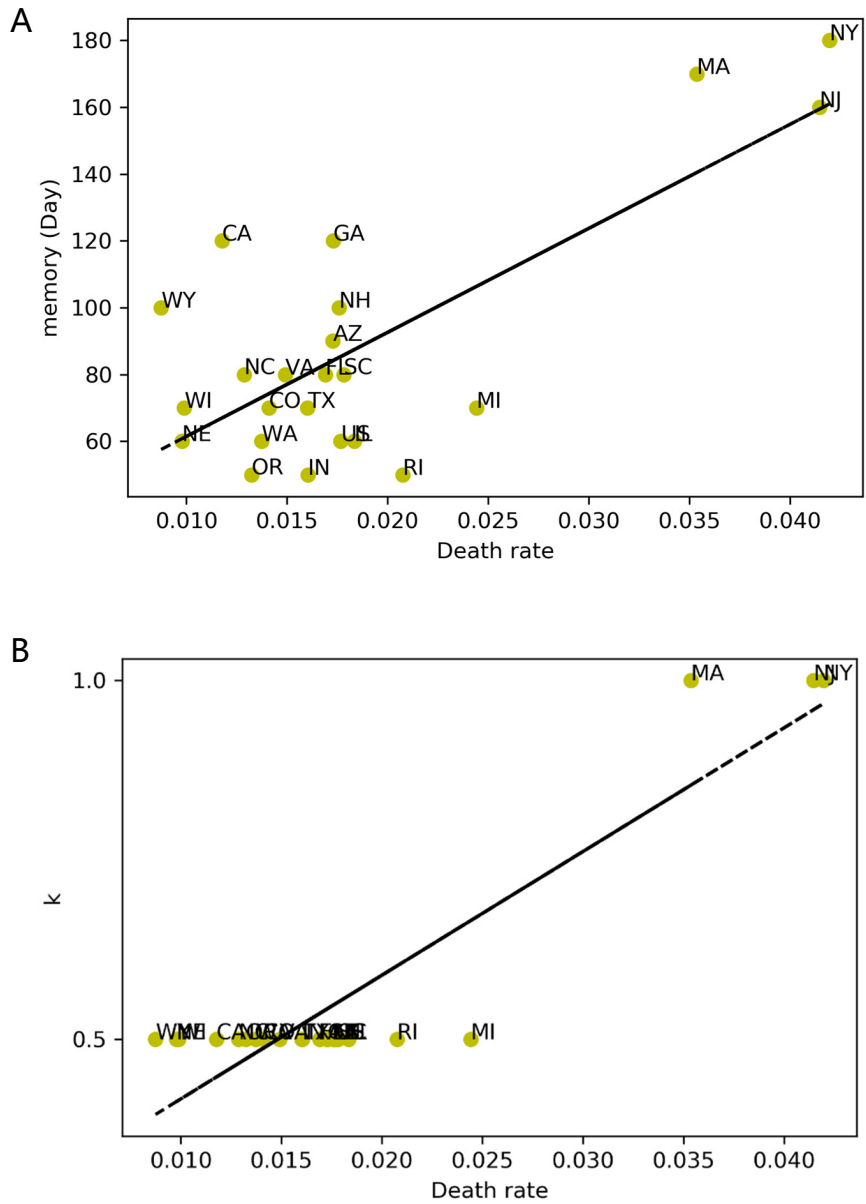

**Figure S1**

(A) Memory length against death rate (total death cases over total positive cases) for United States and its 22 states. The Pearson coefficient is 0.75 with a p-value of $4 \cdot 2 \times 10^{-5}$. (B) Speed of response $k$ against death rate. The Pearson coefficient is 0.91 with a p-value of $1 \cdot 4 \times 10^{-9}$.

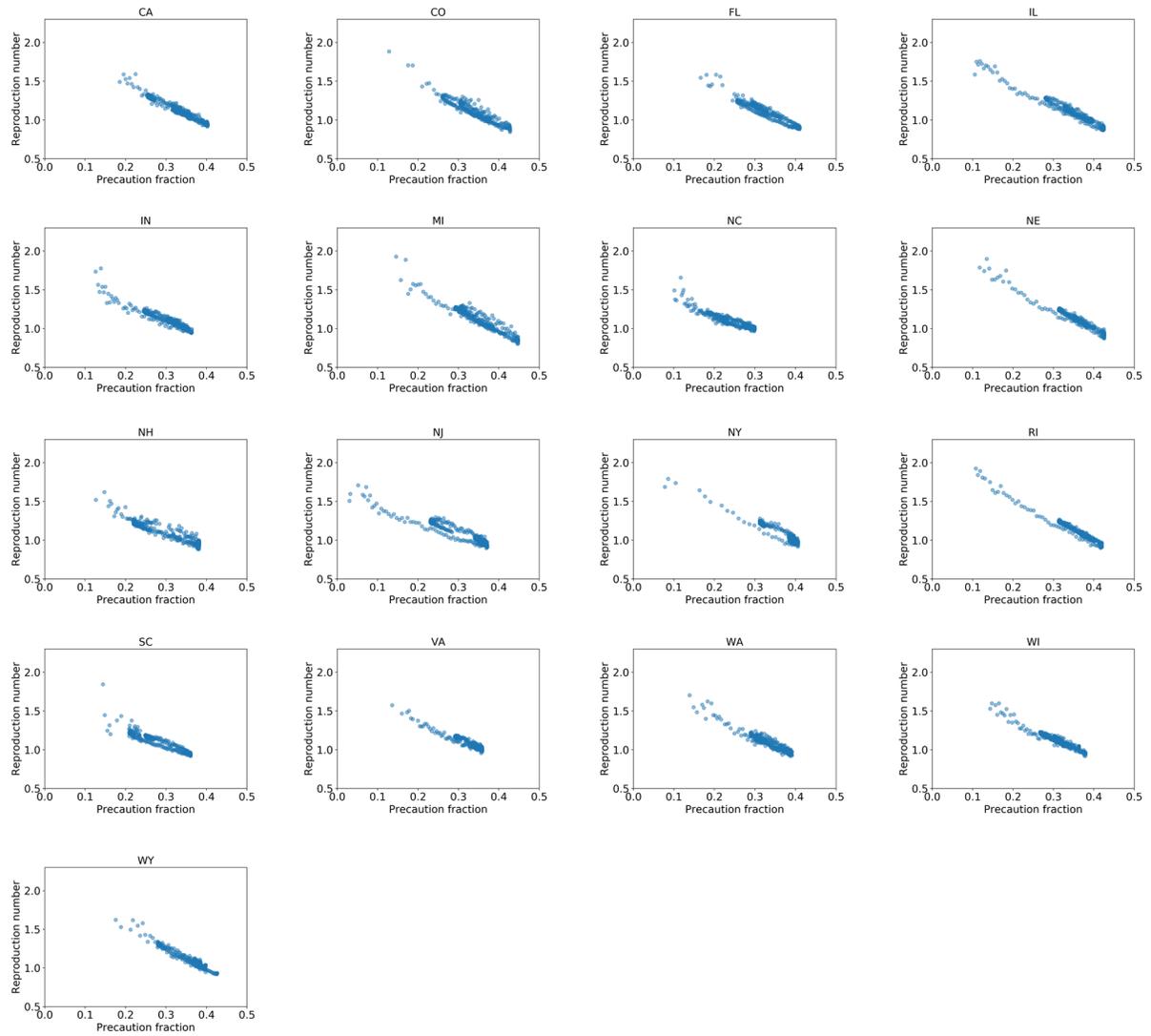

**Figure S2**

Reproduction number against precaution fraction for the remaining states.

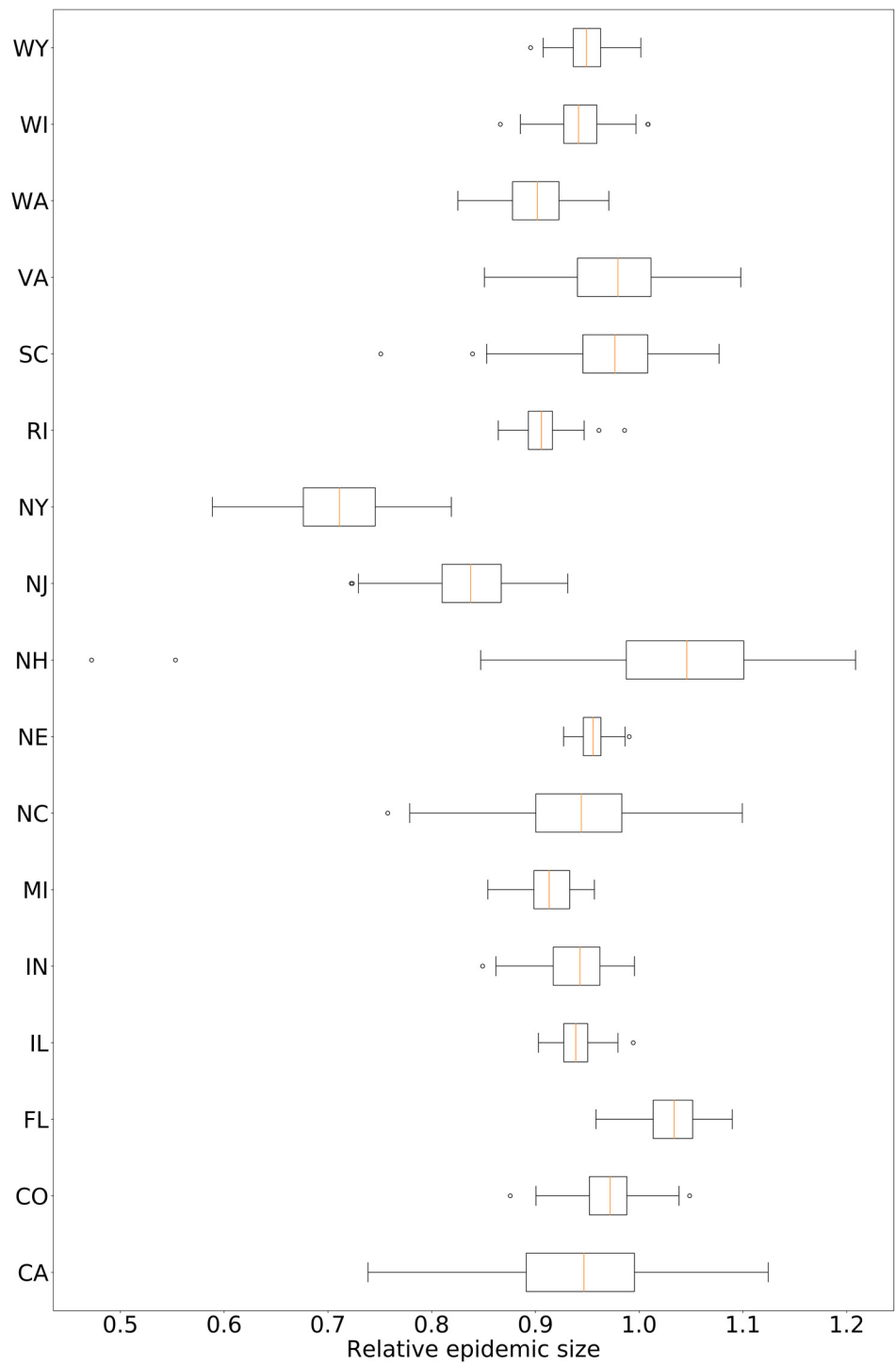

**Figure S3**

Relative epidemic size of 100 model simulations over real data for the remaining data sets.

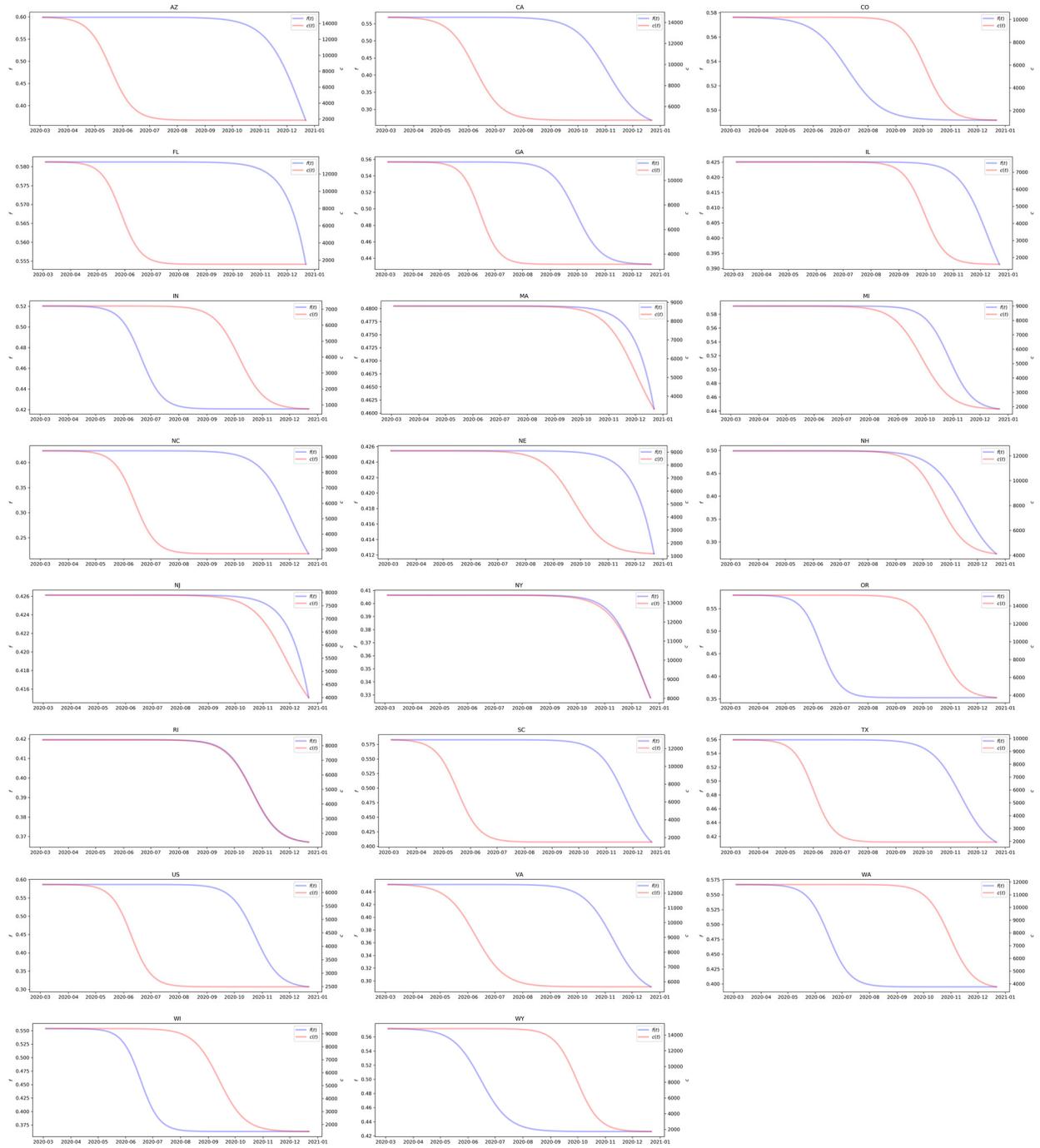

**Figure S4**

Decay functions determined by optimization algorithm in model fitting.

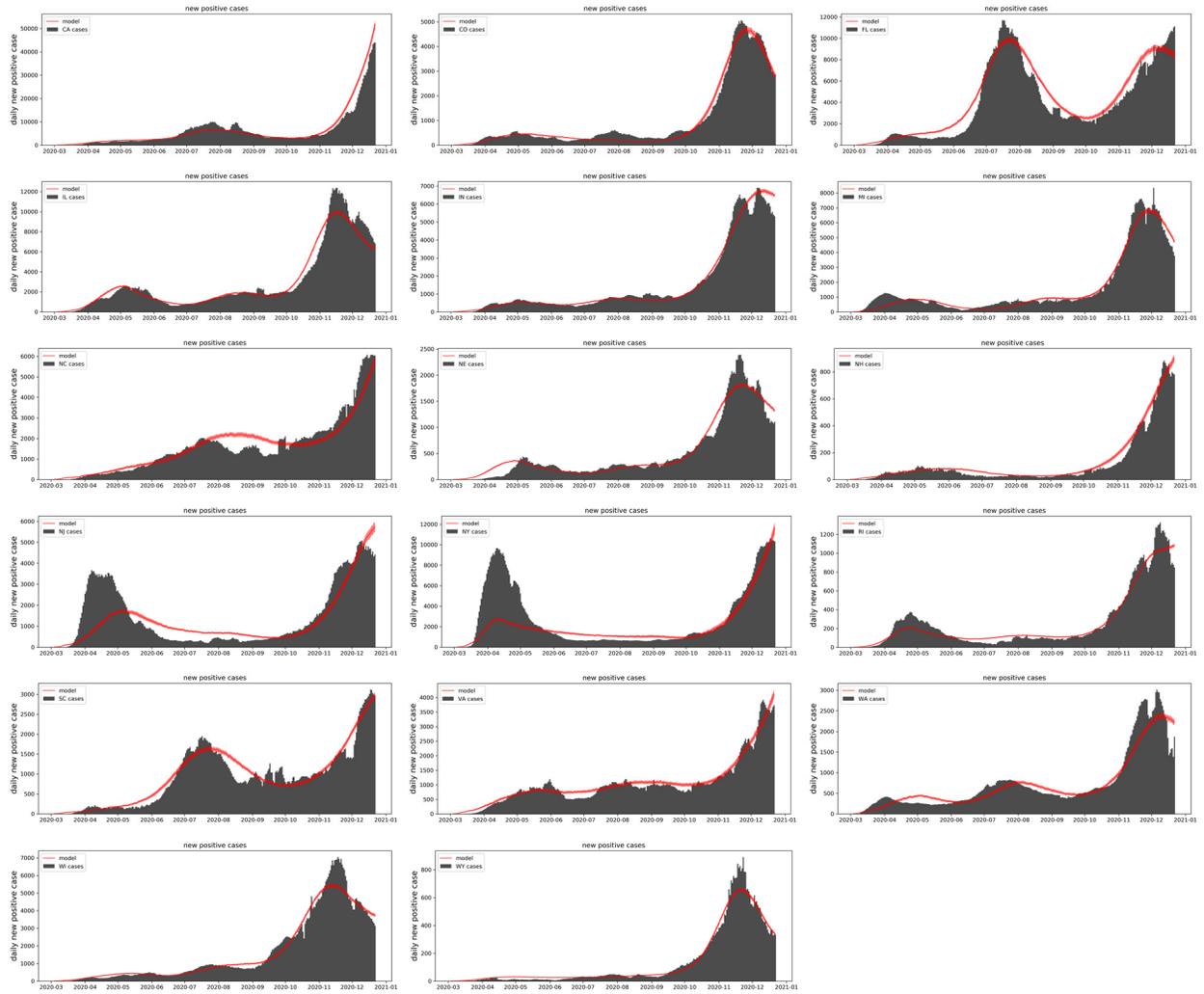

**Figure S5**

Model fitting results for the remaining states.

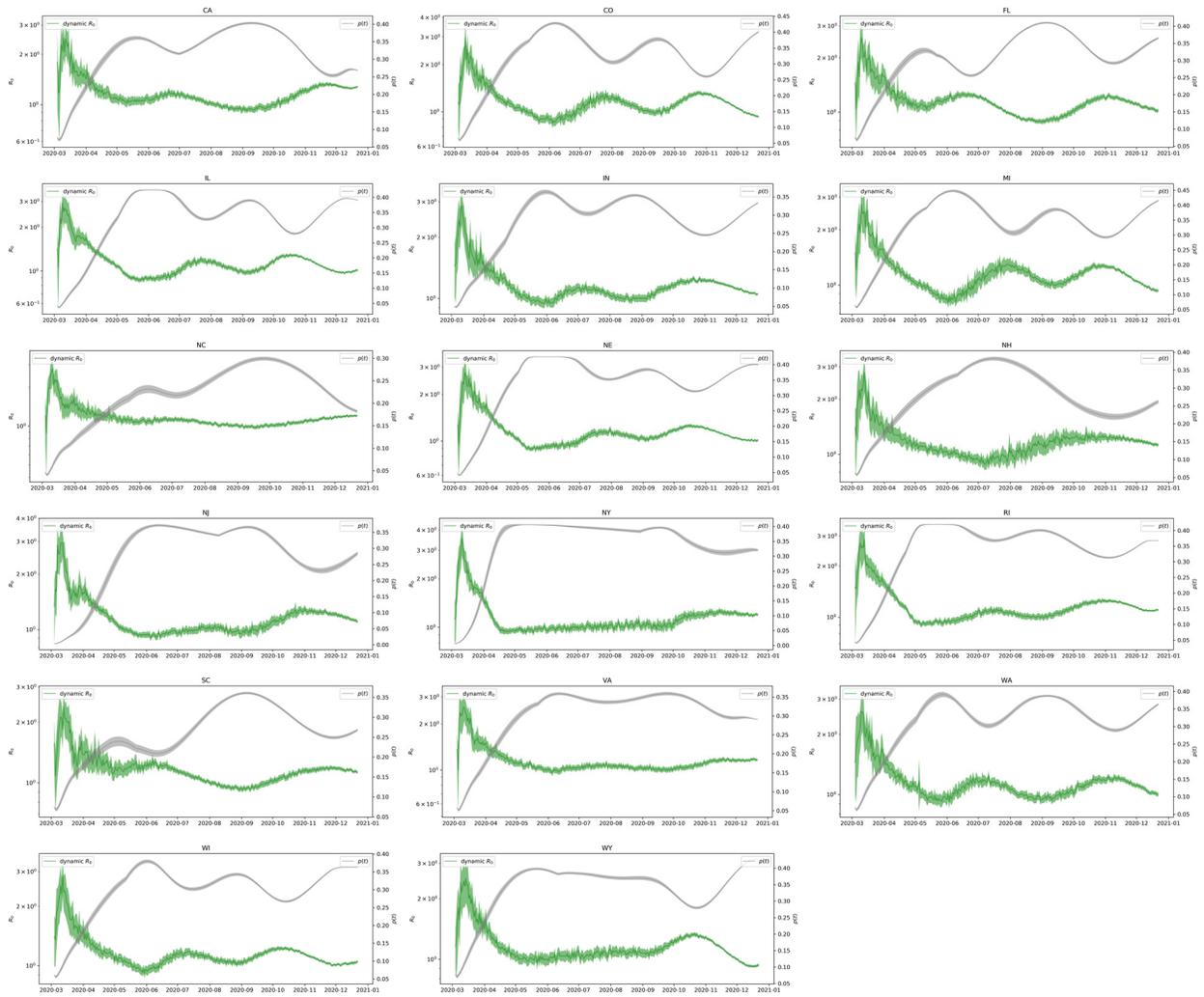

**Figure S6**

Time-varying reproduction number and the vigilance fraction $p(t)$ with their 99% confidence intervals in model fitting results for the remaining states.

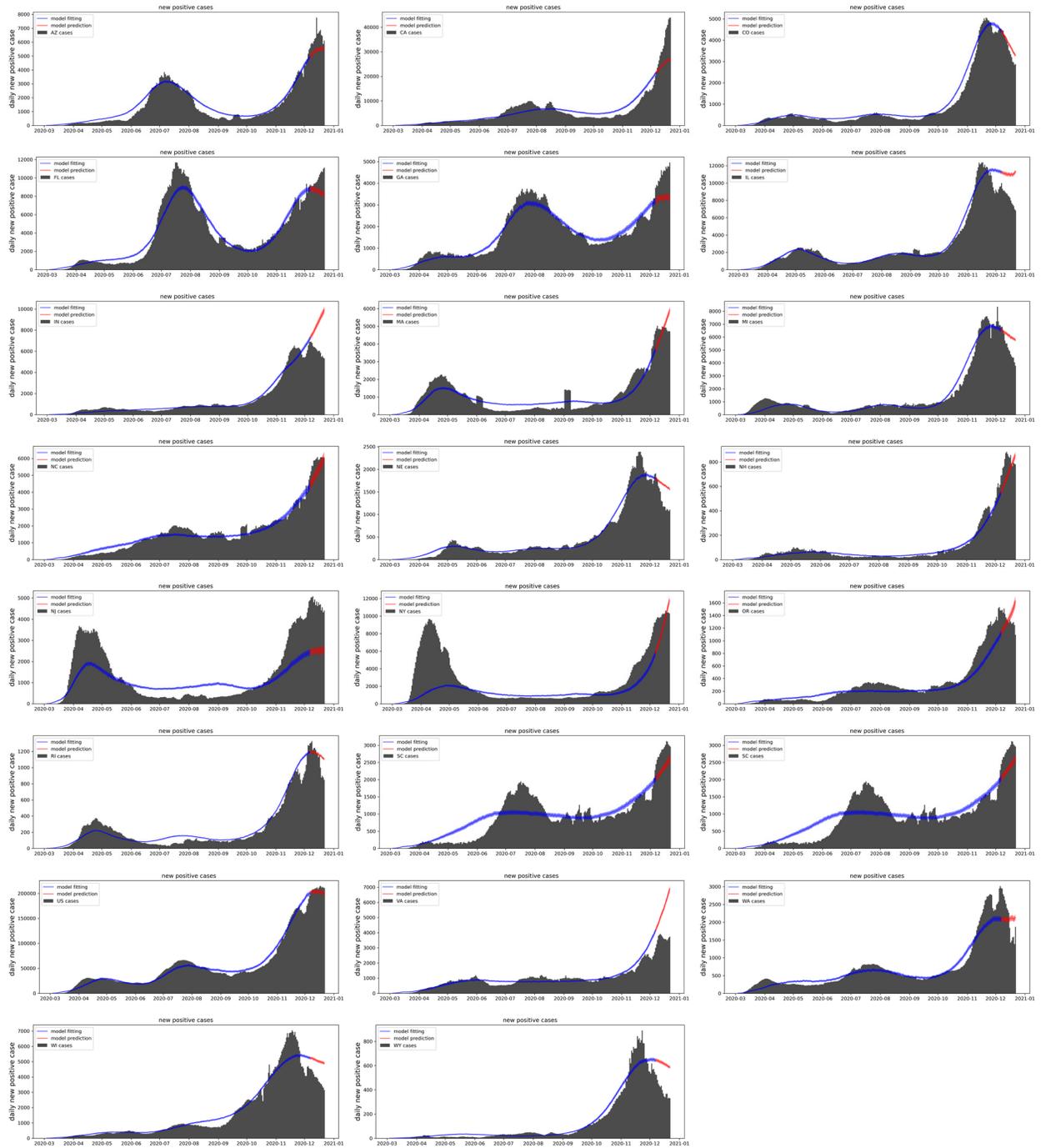

**Figure S7**

15-day prediction results. Bars correspond to the real data, blue line is the average fitting result of 100 simulations with 99% confidence interval (CI) while the red line is the predicted cases of the fitted model.

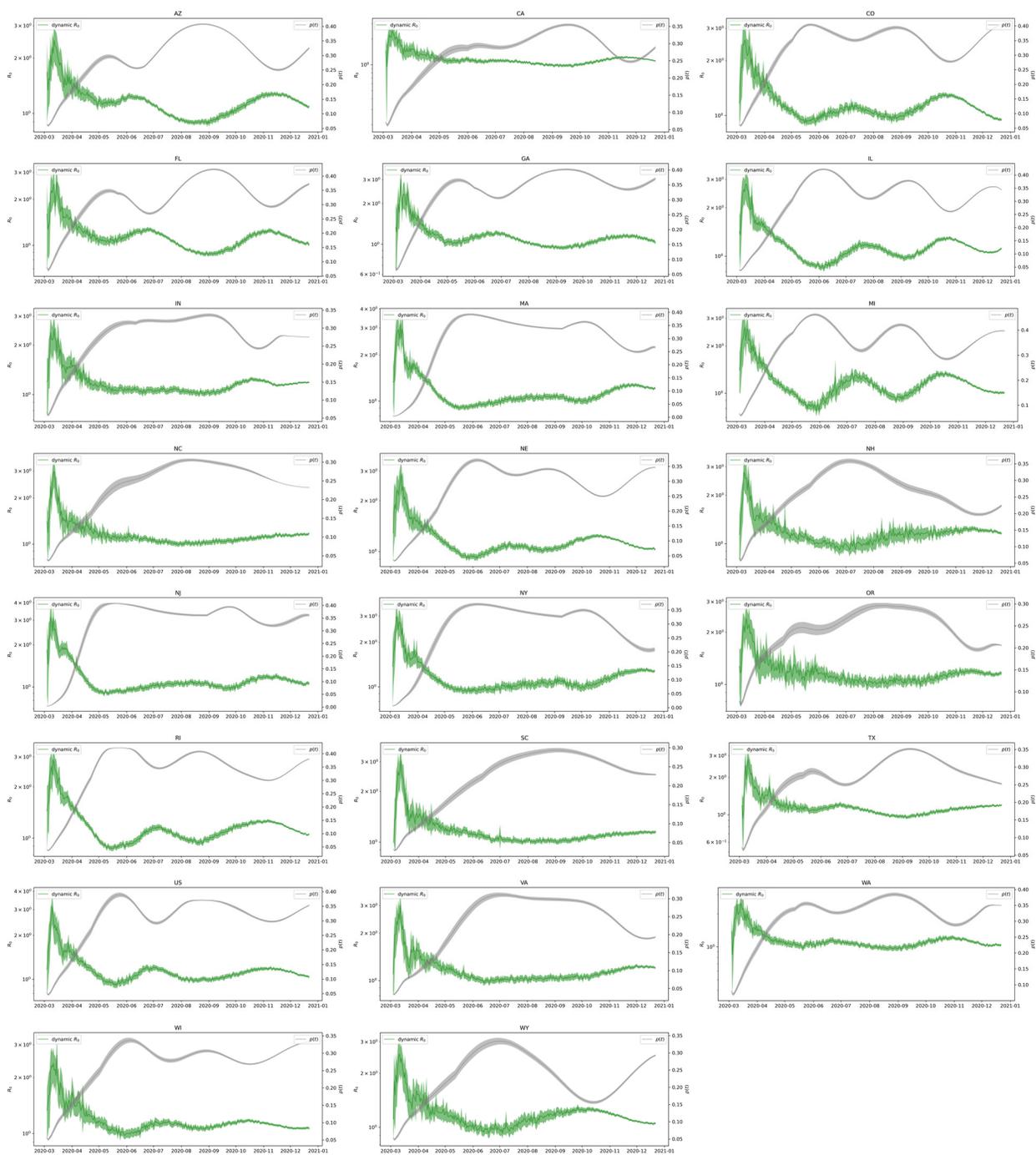

**Figure S8**

Time-varying reproduction number and the vigilance fraction $p(t)$ with their 99% confidence intervals in 15-day prediction results.

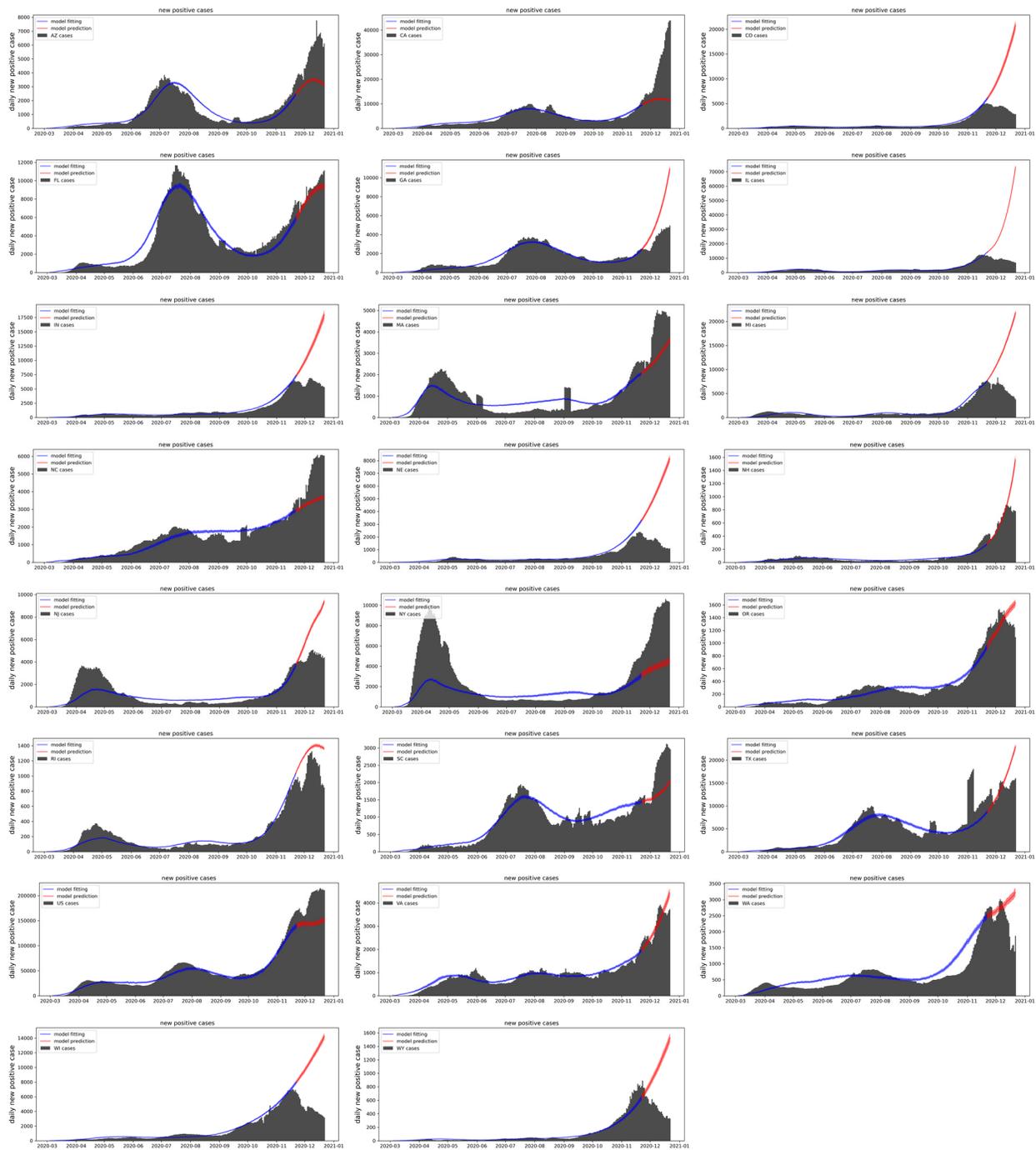

**Figure S9**

30-day prediction results. Bars correspond to the real data, blue line is the average fitting result of 100 simulations with 99% confidence interval (CI) while the red line is the predicted cases of the fitted model.

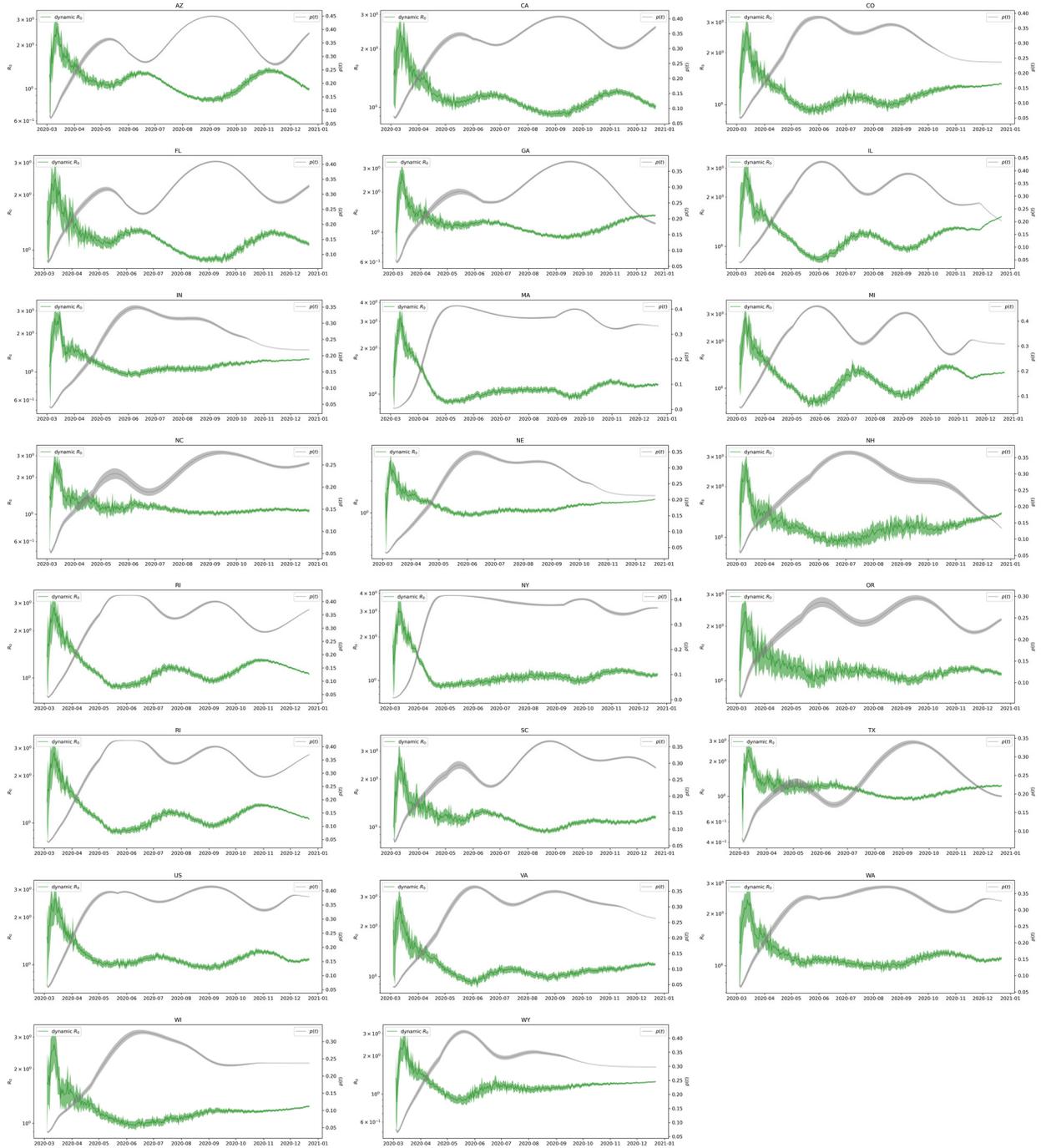

**Figure S10**

Time-varying reproduction number and the vigilance fraction $p(t)$ with their 99% confidence intervals in 30-day prediction results.

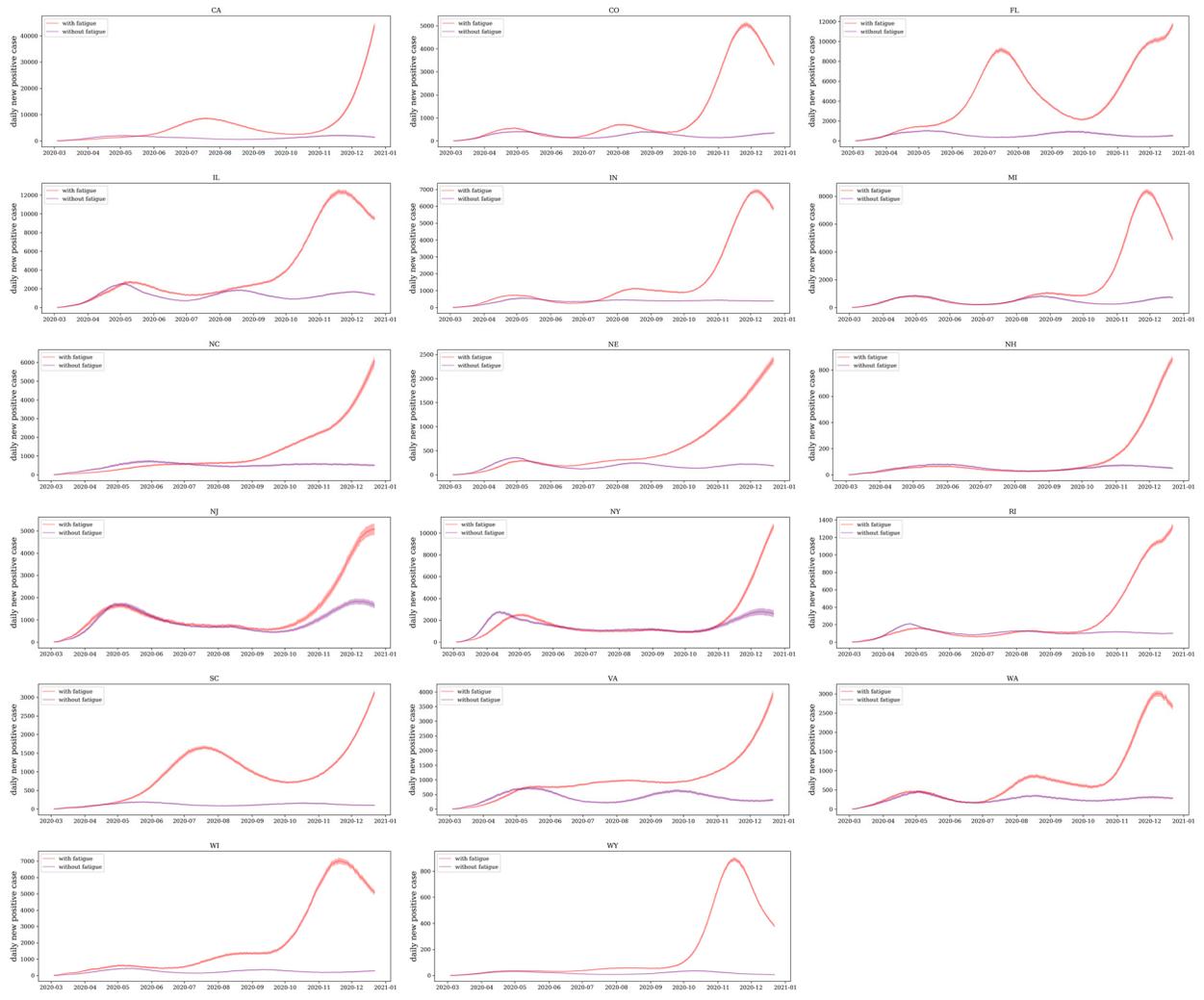

**Figure S11**

Original cases with pandemic fatigue and cases without pandemic fatigue for the remaining states.

**Table S1**

Parameter space of Hyperopt optimization algorithm.

| Parameter | Space | Type |
|---|---|---|
| $m$ | [10, 230] | integer/day (step of 10) |
| $P_0$ | [5, 20] | integer |
| $\xi$ | [0·4, 0·5] | float |
| $f(0)$ | [0·2, 0·6] | float |
| $c(0)$ | [6000, 15000] | integer (step of 100) |
| $l_f$ | [0·1, 0·9] | float |
| $l_c$ | [0·1, 0·9] | float |
| $u_f$ | [6, 18] | float |
| $u_c$ | [6, 18] | float |
| $s_f$ | [10, 20] | float |
| $s_c$ | [10, 20] | float |

**Table S2**

Optimal parameters for fitting.

| Country/State | Memory (Day) | $k$ | $P_0$ | $\xi$ | $f(0)$ | $c(0)$ | $l_f$ | $l_c$ | $u_f$ | $u_c$ | $s_f$ | $s_c$ |
|---|---|---|---|---|---|---|---|---|---|---|---|---|
| US | 60 | 0.5 | 7 | 0.494 | 0.587 | 6300 | 0.518 | 0.395 | 17.558 | 8.808 | 13.343 | 11.071 |
| AZ | 90 | 0.5 | 10 | 0.51 | 0.599 | 14800 | 0.301 | 0.125 | 14.712 | 6.003 | 19.63 | 12.615 |
| CA | 120 | 0.5 | 12 | 0.468 | 0.569 | 14500 | 0.436 | 0.324 | 13.415 | 6.515 | 18.203 | 14.799 |
| CO | 70 | 0.5 | 15 | 0.402 | 0.576 | 10200 | 0.853 | 0.112 | 6.599 | 16.792 | 19.224 | 12.769 |
| FL | 80 | 0.5 | 12 | 0.408 | 0.581 | 13400 | 0.537 | 0.114 | 17.78 | 7.694 | 18.794 | 11.036 |
| GA | 120 | 0.5 | 15 | 0.503 | 0.557 | 11500 | 0.776 | 0.276 | 15.159 | 9.933 | 13.831 | 10.356 |
| IL | 60 | 0.5 | 13 | 0.492 | 0.425 | 7600 | 0.886 | 0.211 | 17.573 | 17.802 | 15.919 | 11.762 |
| IN | 50 | 0.5 | 8 | 0.458 | 0.52 | 7200 | 0.809 | 0.1 | 9.312 | 15.468 | 11.634 | 14.001 |
| MA | 170 | 1 | 10 | 0.434 | 0.48 | 8800 | 0.6 | 0.174 | 17.768 | 15.312 | 18.785 | 17.829 |
| MI | 70 | 0.5 | 16 | 0.449 | 0.592 | 9000 | 0.745 | 0.201 | 17.983 | 13.047 | 13.226 | 15.911 |
| NC | 80 | 0.5 | 5 | 0.408 | 0.423 | 9400 | 0.353 | 0.291 | 15.677 | 8.809 | 17.472 | 11.509 |
| NE | 60 | 0.5 | 17 | 0.493 | 0.425 | 9100 | 0.747 | 0.125 | 16.997 | 11.775 | 19.484 | 17.362 |
| NH | 100 | 0.5 | 8 | 0.409 | 0.499 | 12400 | 0.478 | 0.323 | 13.653 | 15.216 | 18.899 | 15.078 |
| NJ | 160 | 1 | 8 | 0.432 | 0.426 | 7900 | 0.838 | 0.38 | 17.597 | 14.561 | 18.381 | 18.396 |
| NY | 180 | 1 | 16 | 0.48 | 0.406 | 13400 | 0.711 | 0.345 | 17.334 | 15.437 | 16.244 | 18.442 |
| OR | 50 | 0.5 | 9 | 0.481 | 0.58 | 15200 | 0.608 | 0.239 | 8.947 | 16.782 | 10.924 | 13.644 |
| RI | 50 | 0.5 | 17 | 0.497 | 0.42 | 8400 | 0.874 | 0.154 | 16.193 | 15.577 | 14.266 | 14.833 |
| SC | 80 | 0.5 | 6 | 0.473 | 0.583 | 13000 | 0.653 | 0.115 | 17.23 | 6.036 | 15.332 | 12.342 |
| TX | 70 | 0.5 | 8 | 0.503 | 0.56 | 9900 | 0.712 | 0.196 | 15.391 | 8.045 | 16.466 | 11.138 |
| VA | 80 | 0.5 | 10 | 0.43 | 0.452 | 12600 | 0.614 | 0.449 | 15.021 | 6.001 | 16.725 | 16.221 |
| WA | 60 | 0.5 | 11 | 0.436 | 0.567 | 11800 | 0.697 | 0.305 | 8.612 | 17.997 | 12.077 | 13.372 |
| WI | 70 | 0.5 | 10 | 0.434 | 0.554 | 9400 | 0.655 | 0.154 | 10.173 | 13.444 | 10.383 | 14.364 |
| WY | 100 | 0.5 | 16 | 0.484 | 0.572 | 14800 | 0.745 | 0.115 | 6.603 | 16.947 | 15.647 | 12.402 |

**Table S3**

Optimal parameters for 15-day prediction.

| Country/State | Memory (Day) | $k$ | $P_0$ | $\xi$ | $f(0)$ | $c(0)$ | $l_f$ | $l_c$ | $u_f$ | $u_c$ | $s_f$ | $s_c$ |
|---|---|---|---|---|---|---|---|---|---|---|---|---|
| US | 50  | 0.5 | 11 | 0.438 | 0.55  | 7200  | 0.677 | 0.23  | 9.764  | 15.814 | 11.004 | 14.811 |
| AZ | 110 | 0.5 | 9  | 0.47  | 0.478 | 14300 | 0.723 | 0.208 | 11.236 | 7.757  | 19.126 | 10.236 |
| CA | 90  | 0.5 | 7  | 0.482 | 0.588 | 11400 | 0.537 | 0.234 | 15.84  | 6.007  | 15.345 | 16.087 |
| CO | 60  | 0.5 | 16 | 0.419 | 0.586 | 9100  | 0.797 | 0.123 | 7.802  | 16.605 | 10.307 | 13.108 |
| FL | 80  | 0.5 | 13 | 0.425 | 0.592 | 14300 | 0.475 | 0.124 | 17.553 | 7.999  | 18.704 | 11.055 |
| GA | 90  | 0.5 | 16 | 0.407 | 0.585 | 12500 | 0.807 | 0.244 | 8.448  | 8.395  | 18.583 | 11.017 |
| IL | 60  | 0.5 | 11 | 0.469 | 0.433 | 6800  | 0.511 | 0.192 | 17.117 | 17.876 | 17.442 | 11.915 |
| IN | 100 | 0.5 | 8  | 0.447 | 0.583 | 10100 | 0.471 | 0.366 | 17.92  | 6.215  | 12.042 | 16.962 |
| MA | 190 | 1   | 9  | 0.479 | 0.519 | 6200  | 0.383 | 0.845 | 14.016 | 15.753 | 19.242 | 18.567 |
| MI | 60  | 0.5 | 17 | 0.505 | 0.6   | 9500  | 0.768 | 0.1   | 14.659 | 12.698 | 19.82  | 16.332 |
| NC | 50  | 0.5 | 5  | 0.469 | 0.324 | 12600 | 0.711 | 0.535 | 14.832 | 7.092  | 15.799 | 11.843 |
| NE | 50  | 0.5 | 9  | 0.422 | 0.4   | 6600  | 0.875 | 0.169 | 11.798 | 17.316 | 16.652 | 12.348 |
| NH | 80  | 0.5 | 6  | 0.449 | 0.6   | 9200  | 0.636 | 0.122 | 13.271 | 15.006 | 19.783 | 15.431 |
| NJ | 180 | 1   | 13 | 0.468 | 0.411 | 8500  | 0.587 | 0.885 | 16.578 | 16.077 | 18.939 | 15.989 |
| NY | 190 | 1   | 8  | 0.442 | 0.517 | 10800 | 0.266 | 0.66  | 17.995 | 13.492 | 15.621 | 19.837 |
| OR | 50  | 0.5 | 5  | 0.432 | 0.573 | 14800 | 0.342 | 0.392 | 17.019 | 6.191  | 14.24  | 11.776 |
| RI | 50  | 0.5 | 16 | 0.452 | 0.422 | 7100  | 0.759 | 0.128 | 16.886 | 17.988 | 19.281 | 12.706 |
| SC | 100 | 0.5 | 5  | 0.416 | 0.34  | 7300  | 0.643 | 0.545 | 13.613 | 6.006  | 18.73  | 19.099 |
| TX | 70  | 0.5 | 7  | 0.439 | 0.587 | 11800 | 0.35  | 0.146 | 14.122 | 6.002  | 19.994 | 14.196 |
| VA | 50  | 0.5 | 5  | 0.487 | 0.403 | 6200  | 0.472 | 0.241 | 14.623 | 16.144 | 18.315 | 15.426 |
| WA | 70  | 0.5 | 12 | 0.406 | 0.58  | 14900 | 0.598 | 0.397 | 17.333 | 6.994  | 13.478 | 12.405 |
| WI | 60  | 0.5 | 7  | 0.481 | 0.569 | 7600  | 0.59  | 0.148 | 7.078  | 9.877  | 15.609 | 19.481 |
| WY | 70  | 0.5 | 6  | 0.405 | 0.6   | 6000  | 0.499 | 0.175 | 17.276 | 14.513 | 12.321 | 12.74  |

**Table S4**

Optimal parameters for 30-day prediction.

| Country/State | Memory (Day) | $k$ | $P_0$ | $\xi$ | $f(0)$ | $c(0)$ | $l_f$ | $l_c$ | $u_f$ | $u_c$ | $s_f$ | $s_c$ |
|---|---|---|---|---|---|---|---|---|---|---|---|---|
| US | 80 | 0.5 | 19 | 0.467 | 0.572 | 11000 | 0.652 | 0.377 | 11.884 | 6.484 | 19.114 | 13.853 |
| AZ | 90 | 0.5 | 14 | 0.469 | 0.591 | 14300 | 0.385 | 0.142 | 17.586 | 8.471 | 19.823 | 10.189 |
| CA | 90 | 0.5 | 11 | 0.492 | 0.578 | 12600 | 0.836 | 0.263 | 12.796 | 7.899 | 17.888 | 11.696 |
| CO | 50 | 0.5 | 10 | 0.446 | 0.397 | 14300 | 0.594 | 0.899 | 14.966 | 15.644 | 13.869 | 14.223 |
| FL | 90 | 0.5 | 11 | 0.4 | 0.589 | 14400 | 0.318 | 0.113 | 17.976 | 8.327 | 18.922 | 10.011 |
| GA | 100 | 0.5 | 8 | 0.424 | 0.554 | 12900 | 0.13 | 0.167 | 14.372 | 6.477 | 19.695 | 13.384 |
| IL | 60 | 0.5 | 12 | 0.473 | 0.49 | 6000 | 0.343 | 0.283 | 14.561 | 12.267 | 17.612 | 17.779 |
| IN | 50 | 0.5 | 8 | 0.4 | 0.447 | 7400 | 0.485 | 0.42 | 12.536 | 14.86 | 15.823 | 17.002 |
| MA | 180 | 1 | 16 | 0.423 | 0.422 | 8500 | 0.776 | 0.602 | 12.429 | 17.283 | 19.321 | 17.233 |
| MI | 60 | 0.5 | 18 | 0.449 | 0.592 | 7600 | 0.52 | 0.505 | 16.666 | 16.157 | 13.373 | 14.148 |
| NC | 50 | 0.5 | 5 | 0.402 | 0.577 | 8600 | 0.641 | 0.173 | 12.927 | 6.824 | 19.191 | 12.378 |
| NE | 50 | 0.5 | 8 | 0.409 | 0.403 | 6200 | 0.527 | 0.734 | 17.009 | 12.926 | 12.233 | 14.335 |
| NH | 80 | 0.5 | 6 | 0.514 | 0.523 | 12000 | 0.151 | 0.324 | 13.556 | 13.618 | 18.893 | 19.441 |
| NJ | 200 | 1 | 10 | 0.492 | 0.392 | 10200 | 0.643 | 0.503 | 14.692 | 16.027 | 19.006 | 15.417 |
| NY | 190 | 1 | 17 | 0.485 | 0.417 | 13500 | 0.869 | 0.702 | 13.66 | 15.066 | 18.831 | 19.502 |
| OR | 70 | 0.5 | 5 | 0.451 | 0.586 | 13500 | 0.413 | 0.336 | 13.341 | 8.916 | 17.808 | 12.23 |
| RI | 60 | 0.5 | 16 | 0.446 | 0.424 | 8900 | 0.883 | 0.1 | 17.382 | 12.423 | 14.176 | 17.974 |
| SC | 60 | 0.5 | 7 | 0.486 | 0.564 | 12600 | 0.318 | 0.129 | 15.815 | 8.03 | 19.136 | 10.309 |
| TX | 80 | 0.5 | 5 | 0.485 | 0.588 | 14900 | 0.321 | 0.101 | 14.93 | 7.326 | 18.267 | 10.404 |
| VA | 50 | 0.5 | 8 | 0.441 | 0.402 | 9500 | 0.629 | 0.771 | 12.8 | 8.628 | 18.981 | 13.114 |
| WA | 90 | 0.5 | 11 | 0.422 | 0.584 | 12200 | 0.548 | 0.47 | 11.696 | 7.875 | 19.243 | 11.435 |
| WI | 50 | 0.5 | 6 | 0.445 | 0.396 | 7800 | 0.6 | 0.236 | 14.339 | 12.395 | 12.097 | 16.443 |
| WY | 50 | 0.5 | 13 | 0.473 | 0.562 | 14300 | 0.53 | 0.576 | 8.524 | 15.156 | 18.83 | 14.644 |

**Table S5**

Loss and prediction error. It is worth noticing that fitting loss and fitting error are calculated in the same way, they are different because we used 100 simulations for the fitting error and only 1 simulation when fitting to obtain the results faster. The loss of prediction is the relative error on training data and error of prediction is for the test data.

| Country/State | Loss | | | Fitting/Prediction error | | |
| --- | --- | --- | --- | --- | --- | --- |
| | Fitting | 15-day prediction | 30-day prediction | Fitting | 15-day prediction | 30-day prediction |
| US | 0.104137 | 0.116895 | 0.126069 | 0.113338 | 0.037656 | 0.243436 |
| AZ | 0.140132 | 0.189093 | 0.15417 | 0.161838 | 0.158821 | 0.398501 |
| CA | 0.129862 | 0.185445 | 0.139849 | 0.203179 | 0.27438 | 0.541546 |
| CO | 0.116441 | 0.141972 | 0.13189 | 0.12919 | 0.069189 | 2.041213 |
| FL | 0.141804 | 0.148544 | 0.138909 | 0.17923 | 0.143377 | 0.088011 |
| GA | 0.113013 | 0.120552 | 0.13754 | 0.120177 | 0.255078 | 0.603337 |
| IL | 0.120467 | 0.13473 | 0.127611 | 0.15181 | 0.340834 | 2.856132 |
| IN | 0.108787 | 0.144528 | 0.142557 | 0.114343 | 0.426366 | 1.017165 |
| MA | 0.2308 | 0.268249 | 0.23956 | 0.30022 | 0.122239 | 0.308559 |
| MI | 0.124433 | 0.146739 | 0.166666 | 0.156004 | 0.19805 | 1.286533 |
| NC | 0.131524 | 0.130896 | 0.121064 | 0.202857 | 0.10792 | 0.309793 |
| NE | 0.157849 | 0.133618 | 0.117156 | 0.181537 | 0.261076 | 2.4819 |
| NH | 0.153167 | 0.184409 | 0.195939 | 0.181314 | 0.144823 | 0.23976 |
| NJ | 0.28886 | 0.301811 | 0.342541 | 0.397115 | 0.471214 | 0.521886 |
| NY | 0.343875 | 0.41733 | 0.424103 | 0.386519 | 0.174586 | 0.545011 |
| OR | 0.201118 | 0.167437 | 0.175524 | 0.249777 | 0.146319 | 0.157438 |
| RI | 0.171179 | 0.208159 | 0.221826 | 0.195057 | 0.096112 | 0.237852 |
| SC | 0.15877 | 0.184725 | 0.158167 | 0.157753 | 0.190083 | 0.290729 |
| TX | 0.172924 | 0.166649 | 0.173168 | 0.18079 | 0.082429 | 0.198579 |
| VA | 0.147754 | 0.16851 | 0.152001 | 0.123144 | 0.515294 | 0.111637 |
| WA | 0.169515 | 0.154493 | 0.174063 | 0.214338 | 0.237382 | 0.233014 |
| WI | 0.127011 | 0.140602 | 0.124947 | 0.134108 | 0.315959 | 1.511313 |
| WY | 0.156797 | 0.170929 | 0.12378 | 0.146702 | 0.561999 | 1.098091 |

**Table S6**

The reduction in epidemic size (ratio) when there is no precaution fatigue ($l_f = 1$), case fatigue ($l_c = 1$) or pandemic fatigue ($l_f = l_c = 1$).

| Country/State | Precaution fatigue | Case fatigue | Pandemic fatigue |
|---|---|---|---|
| US | 0.427735 | 0.553156 | 0.728053 |
| AZ | 0.197273 | 0.860263 | 0.879869 |
| CA | 0.563515 | 0.514755 | 0.803936 |
| CO | 0.245888 | 0.71931 | 0.773996 |
| FL | 0.010145 | 0.858118 | 0.85911 |
| GA | 0.297297 | 0.591242 | 0.711272 |
| IL | 0.053935 | 0.548937 | 0.557762 |
| IN | 0.314332 | 0.65999 | 0.749428 |
| MA | 0.014687 | 0.358541 | 0.364546 |
| MI | 0.320236 | 0.633773 | 0.702873 |
| NC | 0.193906 | 0.633159 | 0.696993 |
| NE | 0.016793 | 0.641908 | 0.64377 |
| NH | 0.381958 | 0.4809 | 0.63397 |
| NJ | 0.015376 | 0.269503 | 0.274515 |
| NY | 0.108521 | 0.196425 | 0.287904 |
| OR | 0.602456 | 0.343861 | 0.762766 |
| RI | 0.154112 | 0.497906 | 0.568918 |
| SC | 0.152911 | 0.855742 | 0.875037 |
| TX | 0.177496 | 0.764332 | 0.798741 |
| VA | 0.261333 | 0.485733 | 0.61707 |
| WA | 0.457615 | 0.343483 | 0.627648 |
| WI | 0.546159 | 0.680024 | 0.847954 |
| WY | 0.425078 | 0.79714 | 0.869849 |